\documentclass[paper,final,twocolumn,oneside]{IEEEtran}
\usepackage{cite}
\usepackage[dvips]{graphicx}
\usepackage[T1]{fontenc}
\usepackage{dcolumn}
\usepackage{bm}
\usepackage{color}
\usepackage{amssymb}
\newcommand{\figjov}[1]{\includegraphics[width=0.45\textwidth]{#1.EPS}}

\newcommand{\Rfinal}{\mathop{R\mathrm{_{final}}}}
\newcommand{\Rinitial}{\mathop{R\mathrm{_{initial}}}}
\newcommand{\Rhalf}{\mathop{R\mathrm{_{half}}}}
\newcommand{\Imax}{\mathop{I\mathrm{_{max}}}}

\newcommand{\Ipzero}{\mbox{$I\mathrm{_{p}}=0$}}
\newcommand{\Ip}{\mathop{I\mathrm{_{p}}}}
\newcommand{\Ima}[1]{\mbox{$I\mathrm{_{max}}=#1$ mA}}
\newcommand{\T}[1]{\mbox{$T=#1$ K}}
\newcommand{\HM}[1]{\mbox{$H=#1$ Oe}}
\newcommand{\HMzero}{\mbox{$H=0$}}
\newcommand{\CIS}[1]{\mbox{$CIS=#1$\%}}
\newcommand{\shift}[1]{\mbox{$\delta=#1$\%}}
\newcommand{\TMR}[1]{\mbox{$TMR=#1$\%}}
\newcommand{\RAP}{\mathop{R\mathrm{_{AP}}}}
\newcommand{\Rmax}{\mathop{R\mathrm{_{max}}}}
\newcommand{\RP}{\mathop{R\mathrm{_{P}}}}
\newcommand{\Ipz}{\mbox{$I\mathrm{_{p}}=0$}}

\begin{document}

\title{Nanoscopic processes of Current Induced Switching in thin tunnel junctions}
\author{J.~Ventura,
        J.~P.~Ara\'ujo,
        J.~B.~Sousa,
        Y.~Liu,
        Z.~Zhang
        and P.~P.~Freitas,~\IEEEmembership{Member,~IEEE}
\thanks {Work supported in part by POCTI/CTM/59318/2004,
IST-2001-37334 NEXT MRAM and
POCTI/CTM/36489/2000 projects. J. Ventura, Z. Zhang and Y. Liu are
thankful for FCT grants (SFRH/BD/7028/2001, SFRH/BPD/1520/2000 and
SFRH/BPD/9942/2002).}
\thanks{J. Ventura (e-mail: joventur@fc.up.pt), J. P. Araujo
(e-mail: jearaujo@fc.up.pt) and J. B. Sousa (e-mail:
jbsousa@fc.up.pt) are with IFIMUP and Physics Department, Faculty of
Sciences of University of Porto, R. Campo Alegre, 678, 4169-007,
Porto, Portugal.}
\thanks{Y. Liu and Z. Zhang are with INESC Microsystems and Nanotechnologies, R. Alves
Redol 9, 1000-029 Lisbon, Portugal.}
\thanks {P. P. Freitas (e-mail: pfreitas@inesc-mn.pt) is with INESC
Microsystems and Nanotechnologies and also with Instituto Superior
Tecnico, Physics Department, Av. Rovisco Pais, 1000 Lisbon,
Portugal.}} \maketitle

\begin{abstract}
In magnetic nanostructures one usually uses a magnetic field to
commute between two resistance (R) states. A less common but
technologically more interesting alternative to achieve R-switching
is to use an electrical current, preferably of low intensity. Such
Current Induced Switching (CIS) was recently observed in thin
magnetic tunnel junctions, and attributed to electromigration of
atoms into/out of the insulator. Here we study the Current Induced
Switching, electrical resistance, and magnetoresistance of thin
MnIr/CoFe/AlO$_x$/CoFe tunnel junctions. The CIS effect at room
temperature amounts to $6.9\%$ R-change between the high and low
states and is attributed to nanostructural rearrangements of
metallic ions in the electrode/barrier interfaces. After switching
to the low R-state some electro-migrated ions return to their
initial sites through two different energy channels. A low (high)
energy barrier of $\sim$0.13 eV ($\sim$0.85 eV) was estimated. Ionic
electromigration then occurs through two microscopic processes
associated with different types of ions sites/defects. Measurements
under an external magnetic field showed an additional intermediate
R-state due to the simultaneous conjugation of the MR (magnetic) and
CIS (structural) effects.
\end{abstract}


\IEEEpeerreviewmaketitle

\section{Introduction}
\PARstart{M}{agnetic} tunnel junctions (MTJs) consisting of two
ferromagnetic (FM) layers separated by an insulator \cite{Moodera}
are strong candidates for technological applications as non-volatile
magnetic random-access memories (MRAMs) \cite{applications_MRAM2}.
The magnetization of one of the FM layers (pinned layer) is fixed by
an underlying antiferromagnetic (AFM) layer
\cite{Exch_bias_Schuller}, but the magnetization of the other FM
layer (free layer) reverses almost freely when a small magnetic
field is applied. Due to spin dependent tunneling \cite{Tedrow} one
can have two distinct resistance (R) states (the 0 and 1 bits of a
magnetic memory), associated with parallel ($\RP$; low R) or
antiparallel ($\RAP$; high R) pinned/free layers magnetizations.
However, several drawbacks are still of concern for actual MRAM
devices, like cross-talk in the array configuration and large power
consumption, mainly to generate the magnetic field used to commute
R. It is thus desirable to replace the usual magnetic field-driven
by an electrical current-driven resistance switching mechanism. One
such mechanism was predicted by Slonczewski and Berger
\cite{Slow,Berger} and further developed by others
\cite{CIMS_Stiles,CIMS_Zhang} whose studies showed that a spin
polarized current can reverse the magnetization of a FM layer by the
spin transfer effect, as recently observed in nanometer-sized
pillars and exchange-biased spin valves
\cite{CIMS_pillars,CIMS_SVs}, for high current densities
($j\sim10^8$~A/cm$^2$). On the other hand, Liu et al.
\cite{CIS_first} observed R-changes induced by much lower current
densities ($j\sim10^6$~A/cm$^2$) in thin tunnel junctions, which did
not dependent on the relative magnetization orientation in the FM
layers. This new effect was called Current Induced Switching (CIS)
and attributed to electromigration in nanoconstrictions in the
insulating barrier \cite{CIS_Deac}. The underlying physical details
are still poorly understood, though their knowledge can be crucial
to improve device reliability \cite{Diffusion_Reliability_SVs}. One
notices that the CIS and spin transfer \cite{CIMS_TJs,CIMS_TJs2}
effects are likely to coexist in thin MTJs for
$j\gtrsim10^6$~A/cm$^2$. The reasons for the observed dominance of
one effect over the other are still unclear but likely related to
structural differences in the studied TJs.

Here we report a study on the transport properties (electrical
resistance, magnetoresistance and current induced switching) of thin
MnIr/CoFe/AlO$_x$/CoFe tunnel junctions. Current Induced Switching
at room temperature showed a $6.9\%$ resistance change, and the
effect is here discussed in terms of nanostructural rearrangements
of metallic ions from the FM electrodes near the interface with the
insulating barrier. Such rearrangements are mainly reversible and it
is remarkable that more than 10$^4$ R-switching events can be
current-induced without significant damage to the tunnel junction.

Two different electromigration energy barrier channels are observed
($\Delta_1\sim$ 0.13 eV; $\Delta_2\sim$ 0.85 eV), associated with
ion electromigration between different types of sites/defects near
the metal/insulator interfaces, as well as to the lattice binding
energies of such ions. The CIS magnitude decreases with decreasing
temperature (e.g. 3.5\% at \T{120}) showing that the
electromigration effect is temperature assisted. Important
differences were observed in the CIS cycles when measured under a
constant external magnetic field $H$. In particular, one is able to
current-reverse the sign of the exchange bias between the AFM and FM
pinned layer, using an adequate current intensity. The effect is due
to local heating in narrow nanoconstrictions within the oxide
barrier, raising the local temperature above the blocking
temperature of the AFM layer. Thus, in addition to the commonly
observed two R-states, we can obtain a new intermediate R-state,
conjugating the nanostructural and magnetic changes associated
respectively with the CIS and magnetoresistive (MR) effects.

\section{Experimental details}
The complete structure of the Ion Beam Deposited tunnel junction
series \cite{CIS_first} used in this work is glass/bottom lead/Ta
(90 \AA)/NiFe (50 \AA)/MnIr (90 \AA)/CoFe (80 \AA)/AlO$_x$ (3 \AA +
4 \AA)/CoFe (30 \AA)/NiFe (40 \AA)/Ta (30 \AA)/TiW(N) (150 \AA)/top
lead. The AlO$_x$ barrier was formed by two-step deposition and
oxidation processes. NiFe, CoFe, MnIr and TiW(N) stand for
Ni$_{80}$Fe$_{20}$, Co$_{80}$Fe$_{20}$ and Mn$_{78}$Ir$_{22}$,
Ti$_{10}$W$_{90}$(N). The bottom and top leads are made of Al 98.5\%
Si 1\% Cu 0.5\%, 600 \AA\ and 3000 \AA\ thick respectively, and are
26 $\mu$m and 10 $\mu$m wide. The junctions were patterned to a
rectangular shape with areas ($A$) ranging from \mbox{$1\times1$
$\mu$m$^{2}$} to \mbox{$4\times2$ $\mu$m$^{2}$} by a self-aligned
microfabrication process. The samples were annealed at 550 K under
an external magnetic field to impress an exchange bias direction
between the AFM and FM pinned layers, taken here as the positive
direction.

The electrical resistance, magnetoresistance and current induced
switching were measured with a four-point d.c. method and an
automatic control and data acquisition system. Temperature dependent
measurements were performed in a closed cycle cryostat down to 25 K.
The CIS cycles were performed using the pulsed current method
\cite{CIS2}: current pulses ($\Ip$) of \mbox{1 s} duration and
\mbox{5 s} repetition period are applied to the TJ, starting with
increasingly negative pulses from $\Ipzero$ (junction resistance
$\equiv$ $\Rinitial$), in $\Delta I_{p}=2$ or $3$ mA steps until a
negative maximum $-\Imax$ is reached. One then positively increases
the current pulses (with the same $\Delta I_p$), following the
reverse trend through zero current pulse ($\Rhalf$) up to positive
$+\Imax$, and then again to zero ($\Rfinal$), to close the CIS
hysteretic cycle. The junction \emph{remnant} resistance is always
measured in the \mbox{5 s}-waiting periods between consecutive
current pulses, using a low current of 1 mA, thus providing a
$R(I_{p})$ curve for each CIS cycle. This low-current method allows
us to discard non-linear $I(V)$ contributions to the resistance.
Positive electrical current is here considered as flowing from the
bottom to the top lead.

One then defines the CIS coefficient,
\begin{equation}\label{Defenição de CIS}
    CIS =\frac{\Rinitial-\Rhalf}{(\Rinitial+\Rhalf)/2},
\end{equation}
and the resistance shift ($\delta$) in each cycle:
\begin{equation}\label{Defenição de delta}
    \delta
    =\frac{\Rfinal-\Rinitial}{(\Rinitial+\Rfinal)/2}.
\end{equation}
The Tunnel Magnetoresistance is defined as
\begin{equation}\label{Defenição de TMR}
TMR=\frac{\RAP-\RP}{\RP}.
\end{equation}

\section{Experimental results and Discussion}
\subsection{Electrical resistance and CIS effect}
The temperature dependence (300 -- 25 K) of the electrical
resistance of the studied tunnel junction ($R_{300K}\approx5.5~
\Omega$; $R\times A\approx11~\Omega \mu$m$^{2}$ and \TMR{16}) showed
a slight quasi-linear R-increase with decreasing temperature,
indicating a tunnel-dominated behavior (dR/dT<0), although the
presence of few or small pinholes cannot be excluded \cite{Oliver}.
Measured I(V) characteristics (at room temperature) were fitted
using Simmons' model \cite{Simmons_IV} giving a barrier thickness
\mbox{$t=8.3$ \AA}~and a barrier height \mbox{$\varphi=0.7$ eV} (not
shown).

\begin{figure}
\begin{center}
\figjov{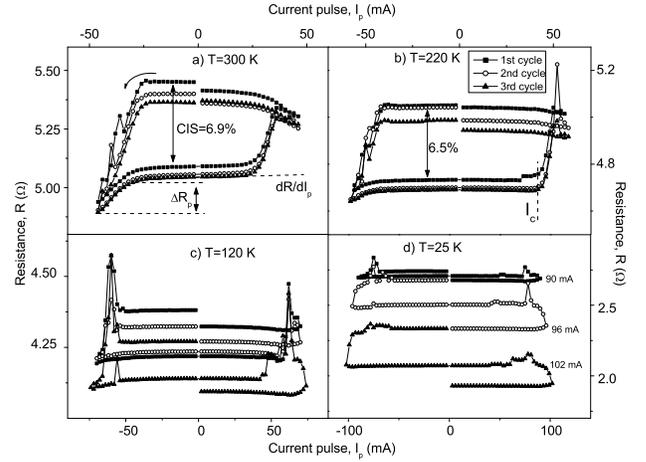}
\end{center}
  \caption{Current Induced Switching cycles at selected temperatures (300 -- 25 K).
  After each current pulse $\Ip$, the electrical resistance of
  the tunnel junction is measured under a low bias current (1 mA).
  R-switching due to electromigration of ions from the electrodes
  into the barrier is observed for $I_p<-I_c$. For $I_p>I_c$ ions return to the
  electrodes and a resistance increase is observed. At low temperatures
  the resistance irreversibly decreases for
  sufficiently high positive and negative current pulses.}\label{fig:CIS_cycles(T)}
\end{figure}

We also performed CIS cycles as a function of temperature in the 300
-- 25 K range, at $\sim$20 K intervals, obtaining the results shown
in Fig. \ref{fig:CIS_cycles(T)}, for some representative curves.
Between such CIS cycles measured at different temperatures, R(T) was
continuously monitored and a negative dR/dT slope was observed.

\begin{figure}
\begin{center}
\figjov{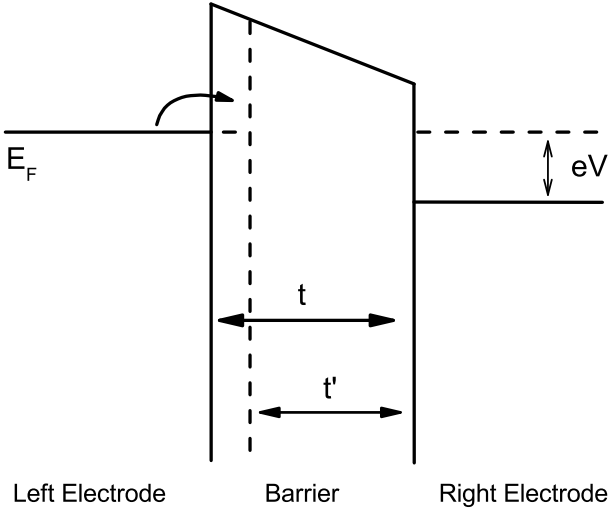}
\end{center}
  \caption{Electromigration-driven barrier thickness decrease
(t$\rightarrow$t'), due to the use of a sufficiently high electrical
current across a thin tunnel junction.}\label{Fig_Fermilevel}
\end{figure}

Figure \ref{fig:CIS_cycles(T)}a) displays three consecutive CIS
cycles measured at \T{300} and using \mbox{$\Imax=46$ mA}. When
increasingly negative current pulses (starting from $\Ipzero$) are
applied, one sees that the TJ resistance remains fairly constant
(\emph{high R-state}) down to $I_{p} \approx-24$ mA (where we define
the critical switching current, $I_c$). At this stage, further
negative increase in pulse intensity, to $-\Ima{-46}$, produces a
sharp resistance decrease (\CIS{6.9}), i.e. switching to a \emph{low
R-state}. This indicates a weakening of the oxide barrier, here
associated with the migration of ions from the metallic electrodes
into the insulator, assisted both by intense electrical fields and
local thermal effects. Notice that even a small barrier weakening
due to such migration (Fig. \ref{Fig_Fermilevel}) could considerably
lower the tunnel resistance due to its exponential dependence on
barrier thickness \cite{Simmons_IV}. The estimated electrical field
at switching is $E\sim 1.5$ MV/cm, considerably smaller than that at
dielectric breakdown in thin tunnel junctions ($\sim 5-10$ MV/cm)
\cite{Oliver3}. On the other hand, local temperatures inside the TJ
can rise above 520 K, as experimentally confirmed below. Such high
temperatures (combined with $E\sim1.5$ MV/cm) are known to be
capable of removing an atom out of its lattice potential well
\cite{heatingmesoscopic}.

Returning to Fig. \ref{fig:CIS_cycles(T)}a, one sees that a slight
increase of the current pulses from $-\Ima{-46}$ up to
$I_p\approx-30$ mA is accompanied by a significant rise in
resistance ($\Delta R_p$). This indicates that the reduction of the
migration driving force (electrical field) allows some atoms to
easily return to their initial sites in the metallic electrodes,
involving low energy barriers ($\Delta_1\sim$ 0.13 eV). However,
most of the displaced ions remain in their local minima inside the
oxide barrier since we still have a low R-state. This indicates that
the displacement of such ions involves higher energy barriers
($\Delta_2\sim$ 0.85 eV; see below).

Only when the current pulse reaches a sufficiently high positive
value ($I_{p} \approx+24$ mA) does electromigration start in the
reverse sense (previously displaced metal ions now move from the
oxide into the electrodes), increasing the effective oxide barrier
and the TJ electrical resistance. Further increase in $I_p$ produces
a small resistance maximum around $+36$ mA. If one reduces the
positive pulses from $+\Ima{46}$ to zero, the plateau of high
constant resistance again emerges ($\Rfinal$) below
\mbox{$\Ip\simeq24$ mA}. The small difference between $\Rfinal$ and
$\Rinitial$ indicates a weak irreversibility, i.e. incomplete tunnel
junction recovery.

A slightly lower CIS effect of 6.5\% (Fig. \ref{fig:CIS_cycles(T)}b)
is observed at \T{220}. A higher $\Imax$ value (\mbox{58 mA}) is
used to achieve R-switching, confirming that the CIS effect is
thermally activated. An interesting feature develops near $+\Imax$
(just after R-switching) leading to a sharp resistance maximum
($\Rmax$). This effect gets more enhanced in the second CIS cycle,
giving an over-resistive state ($\Rmax>\Rinitial$) before declining
to the final resistance.

Notice that R-switching is asymmetric with respect to the applied
current direction. If one starts a CIS cycle with increasingly
positive current pulses, no switching occurs, indicating that only
ions from one electrode/barrier interface are active in
electromigration \cite{CIS_JOV_PRB}. This asymmetry may arise from
the fact that the top electrode is deposited over an oxidized
\emph{smooth} surface, while a much more irregular bottom
electrode/oxide interface is experimentally observed
\cite{Zhang-HRTEM}. Since migration of ions into and out of the
barrier should occur preferentially in localized nanoconstrictions
(where the electrical fields are higher), one concludes that such
ions likely belong to the bottom electrode. In fact, the Current
Induced Switching effect is strongly dependent on the topography of
the electrode/barrier interface as experimentally observed in
magnetic tunnel junctions with different barrier thicknesses and
different insulating barrier materials \cite{CIS_Deac}.

Measurements performed at \T{120} give considerably lower CIS
signals ($\sim$3.5\%; Fig. \ref{fig:CIS_cycles(T)}c), again
requiring higher $\Imax$ (70 mA) for switching. In addition to the
previously observed anomalous resistance maximum near $+\Imax$, a
similar effect also arises near $-\Imax$, again giving an
over-resistive state. In the third CIS cycle (under \Ima{74}), the
resistance suddenly irreversibly decreases when
\mbox{$I_{p}\gtrsim+70$ mA}. This low R-value persists with the
subsequent decrease of $I_{p}$ to zero. Thus, sufficiently high
$I_{p}$ values cause irreversible oxide-barrier degradation.

To ensure resistance switching at \T{25} (Fig.
\ref{fig:CIS_cycles(T)}d) we adopted $\Imax=90$ mA in the first
cycle. The anomalous resistance maxima near $\pm \Imax$ are now much
attenuated but the resistance systematically shows collapsing steps
(irreversible junction degradation) both at positive \emph{and
negative} high current pulses. Subsequent R(T) measurements indicate
metallic conductance in the TJ (dR/dT>0). This means that
metallic-like paths are opened across the insulating barrier
(formation or enlargement of pinholes) while performing CIS cycles
under high $\Imax$ values and now dominate the tunnel junction
conductance.

\begin{figure}
\begin{center}
\figjov{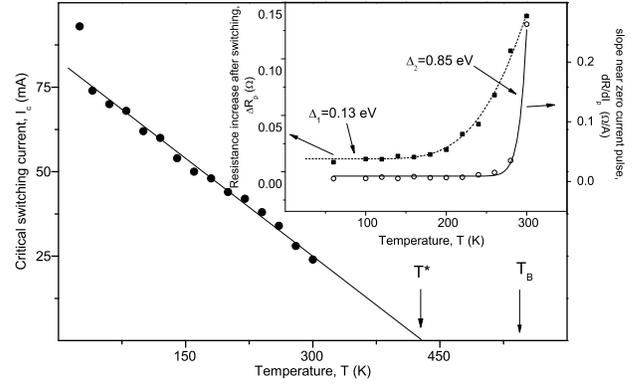}
\end{center}
  \caption{Temperature dependence of the critical current $I_c$ needed to induce resistance switching,
  extrapolating to zero at $T^*\approx425$~K. Inset: temperature
  dependence of the resistance recovery observed after switching (near $-\Imax$; $\Delta R_p$) and of
  the $dR/dI_p$ slope near zero current pulse.}\label{fig7:Ic(T)}
\end{figure}

The temperature dependence of the critical current needed to induce
resistance switching ($I_c$; see Fig. \ref{fig:CIS_cycles(T)}b) was
found to exhibit a quasi-linear decrease with increasing
temperature, extrapolating to zero at $T^*\approx425$ K, as shown in
Fig. \ref{fig7:Ic(T)}. Such behavior can be understood if one
considers the expression for the effective barrier modified by
electromigration \cite{EM_nanobridges} $E_0-\xi I$, where $E_0$ is
the zero-bias electromigration-energy barrier, $\xi$ is a parameter
that measures the change of such activation energy as a result of
the electromigration force \cite{PRB45_9311} and $I$ is the applied
current. Electromigration then occurs when the effective barrier
becomes comparable to the thermal energy. The temperature dependence
of the critical current ($I_c$) is then given by
\cite{EM_gold_contacts}:
\begin{equation}\label{Defenição Icmean}
    I_c\approx\frac{E_0}{\xi}-\frac{k_BT}{\xi},
\end{equation}
where $k_B$ is the Boltzmann constant. Although our data could be
well fitted using this simple model, one should notice that the
effective temperature inside the tunnel junctions is larger than
that at which the measurement takes place (see below), which limits
the quantitative understanding of our results.

On the other hand, the localized resistance increase ($\Delta R_p$)
observed just after switching at high negative current pulses (see
Fig. \ref{fig:CIS_cycles(T)}) shows an exponential temperature
dependence $e^{-\Delta_1/k_BT}$ with $\Delta_1 \approx0.13$ eV
(inset of Fig. \ref{fig7:Ic(T)}). This is attributed to local low
barrier channels for atomic migration of displaced metal ions, from
the barrier into the electrode. Additionally, the slope of the CIS
cycles near $\Ipz$ (in the low R-branch; $dR/dI_p$ in Fig.
\ref{fig:CIS_cycles(T)}a), gives an indication on the remaining high
energy barriers, with $\Delta_2 \approx0.85$ eV. This value is
fairly close to the activation energy for atomic diffusion through
grain boundaries in CoFe/Cu multilayers (\mbox{0.90 eV})
\cite{Energia}. One concludes that ionic electromigration can occur
through two microscopic processes with different energy barriers.
These channels may be associated with electromigration of ions with
different binding energies (and migration energies
\cite{Review_Migration}), or trapped at deep potential sites in the
oxide lattice and/or at oxygen vacancies
\cite{O_vacancies,Al2O3_surface}.

We also studied the influence of the current pulse duration and
current cycling on the CIS effect. The critical switching current
$I_c$ (CIS coefficient) increases (decreases) with shortening pulse
time (down to ns). The CIS coefficient under ns pulses can be
improved by heating the sample up to 400 K. The best results
achieved in these type of structures correspond to a critical
switching current density of 2$\times10^6$~A/cm$^2$ for 10-ns
pulses, leading to a CIS coefficient of 3.8\%. With regard to
current cycling, two effects are observed: first, a temperature
increase leads to a reversible resistance decrease. Then, small
irreversible barrier damage occurs due to the irreversible creation
and/or enlargement of nanoconstrictions and pinholes. For
measurements up to 10$^4$ cycles (using ms-pulses), a stable
junction resistance (indicative of barrier stability) is achieved
after the first 6000 pulses. Once this "equilibrium" resistance is
reached for a certain temperature and current pulse, one can, in
principle, apply much larger pulse numbers.

\subsection{CIS effect under an external magnetic field}
\begin{figure}
\begin{center}
\figjov{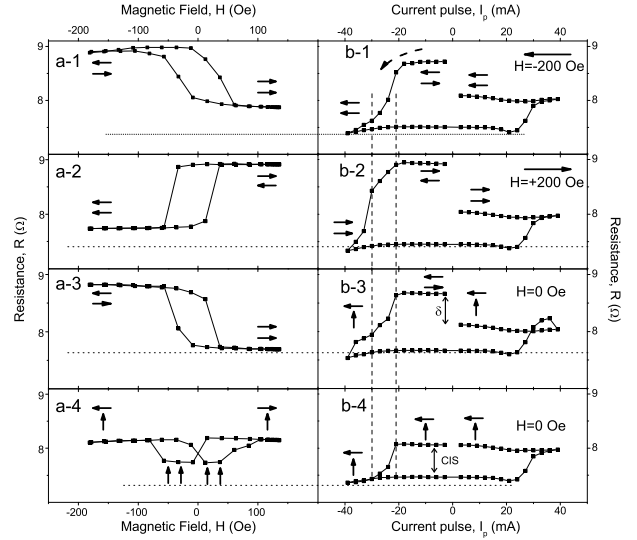}
\end{center}
  \caption{Magnetoresistance (left collum) and subsequent Current Induced Switching
  cycles (right collum) measured under an external magnetic field.
  Notice the inversion of the MR(H)
  behavior after CIS cycles performed under an applied magnetic
  field opposite to the exchange bias.
  Pairs of arrows represent the magnetizations of the pinned (bottom arrow)
  and free layer (top arrow), defining different TJ magnetic states.
  $\uparrow$ denotes a domain state.}\label{fig:CIS(H)}
\end{figure}

A new junction from the same series ($R_{300K}\approx8~\Omega$;
$R\times A\approx48~\Omega \mu$m$^{2}$) was used to study the
influence of the magnetic field on the CIS effect. Figure
\ref{fig:CIS(H)} displays the initial sets of consecutive MR(H) and
CIS (R vs. $I_p$) measurements at room temperature. The CIS cycles
were performed under constant magnetic field of 0 and $\pm 200$ Oe,
and with \Ima{39}.

The MR(H) cycle displayed in Fig. \ref{fig:CIS(H)}a-1 (\TMR{14})
shows the usual low (high) R-state associated with parallel
$\rightrightarrows$ (antiparallel; $\leftrightarrows$) free/pinned
layers magnetizations. The following CIS cycle was measured with
\HM{-200} (Fig. \ref{fig:CIS(H)}b-1), thus starting in the
$\leftrightarrows$ high R-state. One again observes resistance
switching to a \emph{low R-state} for $\Ip\lesssim-21$ mA (\CIS{15};
notice the different denominators in the TMR and CIS definitions).
The final R-value attained at \mbox{-$\Imax$} is lower than the
parallel state resistance $R_P$ of the precedent MR(H) cycle (Fig.
\ref{fig:CIS(H)}a-1). Thus, the observed R-switching cannot be due
to a magnetoresistive effect only, but should also have a structural
contribution.

To clarify the precedent remark one first notices that the
subsequent MR cycle (Fig. \ref{fig:CIS(H)}a-2) appears
\emph{inverted} with respect to that of Fig. \ref{fig:CIS(H)}a-1.
This means that in the previous CIS cycle under \HM{-200} (Fig.
\ref{fig:CIS(H)}b-1) a change in the sign of the exchange bias has
occurred. This effect is attributed to localized heating in
nanoconstrictions in the barrier, under high current pulses, rising
the temperature above the blocking temperature of the MnIr AFM layer
\mbox{($T_B=520$ K; \cite{MnIr_Tb})}. The magnetization of the
pinned layer is then free to align with the applied magnetic field
(\textbf{H} opposite to the initial exchange bias), impressing, upon
cooling below $T_B$, a new (inverted) exchange bias. The junction
magnetic state has then switched from antiparallel
($\leftrightarrows$) to parallel ($\leftleftarrows$) through the
reversal of the pinned layer. The difference between the low R-state
(observed near -$\Imax$) and $R_P$ is then due to electromigration
of metallic ions from the electrodes into the barrier.

Thermally assisted electromigration in the opposite direction
(increasing R) occurs only for $\Ip\gtrsim21$~mA. At the end of the
CIS cycle the junction is in the parallel $\leftleftarrows$ state
and a large difference between $\Rinitial$ and $\Rfinal$
(\shift{-7}) is observed, leading to an \emph{intermediate} R-state.

Although the MR(H) cycle of Fig. \ref{fig:CIS(H)}a-2 is inverted
relatively to that of Fig. \ref{fig:CIS(H)}a-1, the TMR coefficient
remains practically unchanged. The following CIS measurement was
performed under a positive field \HM{+200} (Fig.
\ref{fig:CIS(H)}b-2), giving a CIS coefficient of 18\% and
$\delta=-10$\% (again an intermediate state).

The third MR(H) cycle (Fig. \ref{fig:CIS(H)}a-3) was similar to the
initial one (Fig. \ref{fig:CIS(H)}a-1), indicating that the CIS
cycle under \HM{+200} (Fig. \ref{fig:CIS(H)}b-2) reverted the sample
into its original magnetic state (parallel $\rightrightarrows$
state; the newly induced exchange bias direction is that impressed
during deposition). Subsequent CIS measurements $\emph{without a
magnetic field}$ lead to \CIS{12} and \shift{-7} (Fig.
\ref{fig:CIS(H)}b-3). Notice that when the temperature rises above
$T_B$ the pinned layer magnetization (under $\HMzero$) develops a
complex multi-domain structure (Fig. \ref{fig:CIS(H)}b-3, see
arrows; $\uparrow$ denotes a domain-like state). Accordingly, the
following MR measurement (Fig. \ref{fig:CIS(H)}a-4) shows a double
MR loop with a considerably lower TMR (5\%) due to the lack of full
parallelism between the free and pinned layer magnetizations.
Finally, in the subsequent CIS cycle (Fig. \ref{fig:CIS(H)}a-4),
measured under $\emph{zero magnetic field}$, one has
$\Rfinal\approx\Rinitial$ (resistance shift $\delta=-0.04$\%) and
$CIS=8$\%. The observed R-switching is now only due to the migration
of ions into/out of the barrier (magnetization reversal of the
pinned layer does not occur) and the negligible resistance shift is
attributed to small barrier degradation. Notice that in both CIS
cycles under $H=0$ and $H\neq0$, R-switching always occurs at
\mbox{$\Ip\approx-21$ mA}, indicating that the structural effect
precedes the magnetic one (see corresponding vertical lines in Fig.
\ref{fig:CIS(H)}, column b).

\section{Conclusions} In
conclusion, we presented a detailed study of the Current Induced
Switching effect on low resistance, thin CoFe/AlO$_x$/CoFe tunnel
junctions. We consistently traced the evolution of resistance
switching in consecutive CIS cycles between two (or three) states,
driven by an electrical current, both under $\HMzero$ and $H\neq0$.
Such evolution is controlled by the nanostructural rearrangements of
ions at the electrodes/barrier interfaces (electromigration) and
also by magnetic switching in the pinned layer under sufficiently
high current pulses (under $H\neq0$).

The CIS measurements as a function of temperature (300 -- 25 K)
showed that this effect is thermally assisted. Both low
(\mbox{$\sim$ 0.13 eV}) and high (\mbox{$\sim$ 0.85 eV}) energy
barriers were estimated and associated with electromigration
involving different types of ions sites/defects. At low temperatures
one observes irreversible resistance decreases near $\pm\Imax$,
indicating barrier degradation. If CIS cycles are measured under an
external magnetic field, one is able to current-induce a change in
the sign of the exchange bias of the TJ, and the corresponding
magnetic state (antiparallel to parallel). This effect arises from
excessive local heating in the TJ, and enables us to obtain a CIS
cycle with three different electrical resistance states.

\bibliography{Biblio}

\end{document}